\documentclass{article}

\usepackage{PRIMEarxiv}

\usepackage[utf8]{inputenc} 
\usepackage[T1]{fontenc}    
\usepackage{hyperref}       
\usepackage{url}            
\usepackage{booktabs}       
\usepackage{amsfonts}       
\usepackage{nicefrac}       
\usepackage{microtype}      
\usepackage{lipsum}
\usepackage{fancyhdr}       
\usepackage{graphicx}       
\graphicspath{{media/}}     
\usepackage{amsmath}
\usepackage{float}
\floatstyle{plaintop}
\restylefloat{table}
\DeclareUnicodeCharacter{0307}{\d}

\title{Classification of Gleason Grading in Prostate Cancer Histopathology Images Using Deep Learning Techniques: YOLO, Vision Transformers, and Vision Mamba
}

\author{Amin Malekmohammadi$^{a}$, Ali Badiezadeh$^{b}$, Seyed Mostafa Mirhassani$^c$, Parisa Gifani$^d$, Majid Vafaeezadeh$^{b,}$\thanks{corresponding author}\\\\
$^{a}$ School of Computer Engineering, Iran University of Science and Technology, Tehran, Iran\\
$^{b}$ School of Electrical Engineering, Iran University of Science and Technology, Tehran, Iran\\
$^{c}$ School of Electrical Engineering, Shahrood Branch, Islamic Azad University, Shahrood, Iran\\
$^{d}$ Medical Sciences and Technologies Department, Science and Research Branch, Islamic Azad University, Tehran, Iran\\
}

\begin{document}
\maketitle

\begin{abstract}
\textbf{Purpose:} Prostate cancer ranks among the leading health issues impacting men, with the Gleason scoring system serving as the primary method for diagnosis and prognosis. This system relies on expert pathologists to evaluate samples of prostate tissue and assign a Gleason grade, a task that requires significant time and manual effort. To address this challenge, artificial intelligence (AI) solutions have been explored to automate the grading process. In light of these challenges, this study evaluates and compares the effectiveness of three deep learning methodologies—YOLO, Vision Transformers, and Vision Mamba—in accurately classifying Gleason grades from histopathology images. The goal is to enhance diagnostic precision and efficiency in prostate cancer management.\\
\textbf{Methods:} This study utilized two publicly available datasets, Gleason2019 and SICAPv2, to train and test the performance of YOLO, Vision Transformers, and Vision Mamba models. Each model was assessed based on its ability to classify Gleason grades accurately, considering metrics such as false positive rate, false negative rate, precision, and recall. The study also examined the computational efficiency and applicability of each method in a clinical setting.\\
\textbf{Results:} Vision Mamba demonstrated superior performance across all metrics, achieving high precision and recall rates while minimizing false positives and negatives. YOLO showed promise in terms of speed and efficiency, particularly beneficial for real-time analysis. Vision Transformers excelled in capturing long-range dependencies within images, although they presented higher computational complexity compared to the other models.\\
\textbf{Conclusions:} Vision Mamba emerges as the most effective model for Gleason grade classification in histopathology images, offering a balance between accuracy and computational efficiency. Its integration into diagnostic workflows could significantly enhance the precision of prostate cancer diagnosis and treatment planning. Further research is warranted to optimize model parameters and explore the applicability of these deep learning techniques in broader clinical contexts. The code is available at https://github.com/Swiman/mamba-medical-classification.
\end{abstract}


\section{Introduction}
Gleason cancer, a type of prostate cancer, is characterized by its aggressiveness and progression rate, making precise diagnosis and grading imperative for effective treatment planning \cite{gleason1992histologic}. The current gold standard for diagnosing Gleason cancer relies heavily on histopathological examination of prostate tissue samples, a process that demands considerable expertise, involves lengthy procedures, and may be subject to inconsistencies in interpretation among medical professionals \cite{dominguez2024systematic}. This has led to a pressing need for automated, objective, and reliable diagnostic tools that can complement or even surpass the capabilities of human pathologists \cite{zhu2024harnessing}. The classification of histopathology images for Gleason cancer stands out for its potential to significantly enhance diagnostic accuracy and patient outcomes \cite{bernardino2024using}.\\ 
The advent of artificial intelligence (AI) has revolutionized the diagnosis and prognosis of prostate cancer \cite{niazi2016visually}. Early machine learning techniques for prostate carcinomas were built around manually crafted feature extraction, followed by feature selection, and concluding with traditional classification algorithms \cite{iqbal2021prostate}. However, advancements in technology have led to the emergence of deep learning systems, which utilize multi-layered neural networks as an innovative approach to these traditional feature-based methods for medical image analysis in any modalities \cite{gifani2021automated, gifani2023automatic, shalbaf2022automatic, vafaeezadeh2021deep}. Unlike their predecessors, deep learning techniques do not rely on pre-defined features but can autonomously identify and learn intricate, relevant features straight from the data \cite{vafaeezadeh2022automatic}. This capability mirrors the human brain's ability to process information and generate decision-making patterns. Innovations in neural network design and training methodologies have unlocked the potential to tackle complex learning challenges that were once deemed unsolvable. Consequently, recent studies have concentrated on leveraging deep learning as the cutting-edge technique in machine learning.\\ 
YOLO (You Only Look Once) \cite{redmon2016you} has emerged as a pivotal model in the realm of real-time object detection, offering a unique blend of speed and accuracy that distinguishes it from many other models. Originally designed for general object detection tasks, YOLO's architecture has been adapted to meet the specific demands of histopathology image classification \cite{al2022deep}, where rapid analysis is often as critical as the accuracy of the diagnosis itself. By leveraging YOLO's ability to process images in a single pass, researchers and clinicians can significantly accelerate the detection of pathological features associated with Gleason's cancer, thereby facilitating more timely and accurate diagnoses.\\
Transformers, initially designed for natural language processing tasks, have revolutionized the field of computer vision and medical image analysis \cite{vafaeezadeh2023carpnet, vafaeezadeh2024ultrasound}, including histopathology image analysis \cite{xu2023vision}. These models, inspired by the mechanism of attention in neural networks, excel in capturing complex relationships between elements within an image. By adapting transformers for image classification in histopathology, researchers aim to leverage their ability to focus on relevant features and ignore irrelevant ones, potentially leading to more accurate and nuanced diagnoses of conditions like Gleason's cancer.\\
In the quest to enhance the analytical capabilities of histopathology image analysis, Mamba stands out as a cutting-edge tool \cite{nasiri2024vim4path, ding2024combining, yang2024mambamil}. Originating from the realm of time series analysis, Mamba offers a sophisticated approach to modeling dynamic data, which is particularly pertinent in the context of histopathology where changes over time can significantly influence disease progression and diagnosis. This section delves into the application of Mamba in histopathology image analysis, highlighting its unique ability to capture and analyze temporal patterns within images. By focusing on Mamba's potential to handle dynamic data, we explore how this state-of-the-art method can contribute to more insightful and accurate interpretations of histopathological images, ultimately aiding in the early detection and management of diseases such as Gleason's cancer.\\
In this article, we explore three cutting-edge architectures for classifying Gleason cancer images from Tissue Microarray (TMA), prostate samples: the YOLO structure, Vision Transformer structure, and state-space model (Mamba) structures. These methods, rooted in the latest advancements in deep network technologies, showcase unparalleled efficiency and promise in dissecting the complexities inherent in histopathology images. By harnessing the power of these state-of-the-art techniques, we aim to illuminate the potential of employing deep networks in enhancing the accuracy and reliability of Gleason cancer diagnoses. Consequently, this article is structured to explore each method, offering an overview of their applications, benefits, and limitations within the context of histopathology image analysis.\\
Our article is divided into five sections. In the Literature Review section, we review the most recent works on Gleason grading classification based deep learning methods. The Materials section, we present the datasets used in our study. In the Methods section, we discuss the deep learning methodologies YOLO, Vision Transformers, and Vision Mamba, explaining their appropriateness for this task. In In the results section, we reveal the performance of each method in diagnosing Gleason's cancer. This aids in understanding the advantages and disadvantages of each deep learning method. In the discussion section, we examine the implications of our findings, evaluating the pros and cons of these methodologies in histopathological analysis. Finally, in the conclusion section, we aim to provide a comprehensive summary of our investigative process, concluding with our findings.

\section{Literature Review}
A review study introduced \cite{zhu2024harnessing} the significant role of AI in enhancing prostate cancer diagnosis and treatment through digital pathology and whole-slide imaging. It emphasized the potential of AI to collaborate with pathologists, reducing workload and aiding in treatment decisions. Additionally, the paper outlined the development process and challenges of AI pathology models for prostate cancer, providing access to public datasets and open-source codes to foster research advancements. In \cite{bernardino2024using} the researchers conducted a systematic literature review using specific keywords in major databases to gather relevant information on prostate cancer, active surveillance, and biomarkers. A systematic comparison has been proposed in \cite{dominguez2024systematic} wherein various learning approaches on heterogeneous datasets for Gleason grading and scoring, achieved high performance and generalization. The dataset included nine heterogeneous datasets, allowing for a comprehensive evaluation of the methods. The results showed that fully-supervised learning excelled in patch-level classification tasks, while Multiple-Instance Learning (MIL) methods, particularly Clustering-constrained Attention Multiple Instance Learning (CLAM) \cite{lu2021data}, achieved the highest performance in image-level classification tasks. \\
Arvaniti et al.'s \cite{arvaniti2018automated} study introduces a deep learning method for automating Gleason grading of prostate cancer tissue microarrays through a convolutional neural network (CNN) trained on 641 patients and validated on another 245 patients. The model demonstrated inter-annotator agreement akin to human pathologists, suggesting its potential for enhancing objectivity and reproducibility in prostate cancer grading. It also showed promise in stratifying patients into distinct prognostic groups using disease-specific survival data. Nonetheless, the study identified limitations, including occasional misclassification in unrecognized stromal areas, highlighting the need for additional preprocessing steps to exclude such regions. The inherent subjectivity of the Gleason scoring system and potential discrepancies among pathologists from different institutions underscore the requirement for broader, multicenter studies to improve the model's reliability and applicability. \\
In \cite{rabbani2023unsupervised} the authors proposed Unsupervised Confidence Approximation (UCA) to address training with noisy labels and selective prediction simultaneously. They evaluated the efficacy of UCA on CheXpert and Gleason-2019 datasets \cite{nir2018automatic}, showing performance improvements in both noisy label training and selective prediction aspects. Their method's notable strength is that it provides concurrent solutions for training with noisy labels and selective prediction.
The approach developed by the researchers in \cite{muller2024deepgleason} employs an accessible AI-powered image recognition framework called DeepGleason, which utilizes a tile-by-tile classification strategy incorporating a ConvNeXt architecture and sophisticated image refinement techniques. This system was developed and tested using a proprietary prostate cancer database consisting of 34,264 labeled tiles derived from 369 prostate carcinoma specimens. The data source comprised comprehensive histopathological images obtained from prostate tissue samples. The outcomes demonstrated that DeepGleason attained a macro-averaged F1-score of 0.806, area under the curve (AUC) of 0.991, and accuracy rate of 0.974. In comparison to other architectures, ConvNeXt exhibited superior performance on this particular dataset. The primary strength of this method lies in its exceptional precision and dependability in Gleason grading assessments.
Overcoming the limitations posed by insufficient and unbalanced databases during model training, the method proposed in \cite{golfe2023progleason} presents a conditional Progressive Growing GAN framework named ProGleason-GAN for synthesizing prostate tissue patches with any Gleason Grade. The dataset utilized is SICAPv2 \cite{silva2020going}, and the results indicate that the proposed ProGleason-GAN framework achieved a weighted Frechet Inception Distance (FID) metric of 77.85 for all Gleason grades. A notable advantage of this method is its capacity to generate realistic prostate tissue patches with varying Gleason Grades.\\
In \cite{gifani2024transfer} the authors utilized transfer learning with pre-trained CNNs on Gleason-2019 datasets images, achieving the best performance with the NasnetLarge architecture with an accuracy of 0.93 and an area under the curve of 0.98. Their method's salient feature is the ability to automate Gleason grading effectively.\\
Lucas et in \cite{lucas2019deep} proposed a method for detection of prostate biopsies. Their method involved re-training a CNN for Gleason pattern detection in prostate biopsies, achieving high accuracy in differentiating between Gleason patterns, with a dataset of 96 biopsies from 38 patients. The results demonstrated 92\% accuracy in distinguishing non-atypical from malignant areas (GP $\geq$ 3) and 90\% accuracy in differentiating between GP $\geq$ 4 and GP $\leq$ 3. The method's good point is the high accuracy, sensitivity, and specificity in Gleason pattern classification.\\
In \cite{cserbuanescu2020agreement} the authors used transfer learning from AlexNet and GoogleNet on a dataset from the Gleason-2019 Challenge to classify 6000 cropped images, achieving 85.51\% accuracy for AlexNet and 74.75\% for GoogleNet when compared against the majority vote of pathologists. A hallmark of their method lies in successful classification using pre-trained deep-learning networks.\\
in \cite{niazi2016visually} a method has been proposed based on visually meaningful features to differentiate between low and high-grade prostate cancer, achieving high training and testing accuracies of 93.0\% and 97.6\%, respectively. The dataset consisted of 43 ROI images from TCGA for training and 88 ROIs from an independent dataset for testing. The method's key strengths are its ability to achieve expert-level accuracy in prostate cancer diagnosis and prognosis tasks, while also providing visually interpretable features that can assist pathologists in understanding the AI's decision-making process.\\
A recent investigation \cite{marron2023comparative} centered on examining discrepancies in interpretations among medical experts using a localized collection of 80 comprehensive histopathology images marked by specialists and developing convolutional neural network architectures based on this dataset. The researchers utilized a localized collection of 80 comprehensive histopathology images marked by specialists. The results showed an inter-observer variability of 0.6946 with a 46\% discrepancy in area size of annotations by pathologists, while the best trained models achieved a performance of 0.826±0.014 on the test set. The method's standout feature is its successful demonstration of how AI-driven approaches can potentially mitigate inconsistencies in diagnoses among specialists.
Fabian León and Fabio Martínez \cite{leon2022multitask} introduced a novel approach to address the challenges of Gleason score classification in HARVARD Data-verse \cite{arvaniti2018automated}, comprising 886 hematoxylin and eosin-stained tissue samples, leveraging a multitask deep learning framework. This method combines a triplet loss scheme with a cross-entropy task to create a robust embedding representation that captures the high inter- and intra-class variability inherent in Gleason scoring. The proposed approach significantly improves upon existing methods by attaining an average accuracy of 66\% and 64\% across evaluations by two medical specialists, with no statistical difference, and an impressive 73\% accuracy in patches where both specialists agreed. Despite these strengths, the method still faces limitations, particularly in capturing detailed variations in annotator segmentations, indicating potential areas for future improvement to enhance its effectiveness.
In another research work \cite{qiu2022automatic} the authors proposed a PSPNet for automatic prostate Gleason scoring, utilizing a dataset of 321 biopsy samples from the Vancouver Prostate Centre, achieving top ranking in the Gleason-2019 prostate segmentation benchmark. Their method excelled in distinguishing benign from malignant and high-risk (Gleason pattern 4, 5) from low-risk (Gleason pattern 3) cancer. One of the notable strengths of this technique lies in its ability to excel in both image segmentation and classification tasks, demonstrating its versatility and effectiveness across different medical image analysis domains. Additionally, the method's effectiveness in extracting features from digital histopathology images, combined with its robustness and generalizability across high-resolution images and varying cell and tissue heterogeneities, positions it as a valuable tool in prostate cancer diagnosis. However, the paper also identifies key weaknesses, such as the method's dependency on high-quality expert annotations and challenges in differentiating specific Gleason grades, highlighting areas for further improvement and integration with clinical practices.\\
In other study \cite{kong2024federated}, a novel federated learning framework, Federated Attention Consistent Learning (FACL) framework was used with a dataset of 19,461 whole-slide images of prostate cancer from multiple centers. FACL achieved an AUC of 0.9718 in the diagnosis task and a Kappa score of 0.8463 in the Gleason grading task, outperforming multiple centers. The method's good point lies in its ability to maintain high accuracy and robustness across large-scale, multicenter pathological image datasets while preserving patient privacy. Its weaknesses include the complexity of implementing attention consistency and differential privacy, which might require advanced technical expertise.\\
The field of artificial intelligence (AI) has witnessed remarkable advancements in recent years, significantly impacting various domains, including healthcare. Within the realm of histopathology image analysis, AI has emerged as a powerful tool for enhancing diagnostic accuracy and efficiency. While YOLO, Transformers, and Mamba represent significant strides in applying AI to histopathology, the broader landscape of AI methods extends far beyond these innovations. The next paragraphs aim to illuminate the diversity of AI techniques available for histopathology image analysis, encompassing a range of approaches from the latest deep learning architectures. By exploring the full spectrum of AI methods applicable to histopathology, we seek to underscore the multifaceted nature of AI's contribution to this critical aspect of medical diagnostics, paving the way for further innovation and refinement in disease detection and characterization.

\section{Materials}
Our study leverages two comprehensive datasets, namely the Gleason 2019 dataset \cite{nir2018automatic} and the SICAPv2 dataset \cite{silva2020going}, to explore the nuances of prostate cancer tissue imagery. Each dataset offers unique insights, yet they come with their own set of limitations that necessitate careful consideration during analysis.\\
The study identifies five primary repositories for prostate cancer tissue imagery, with the Cancer Genome Atlas initiative holding the largest collection of approximately 720 biopsy slides \cite{weinstein2013cancer}. However, the lack of Gleason grade annotations at both slide and biopsy levels limits these datasets' applicability \cite{silva2020going}. In contrast, the database by Arvaniti et al. \cite{arvaniti2018automated} offers detailed pixel-level annotations for Gleason patterns across 886 smaller slide segments, though these do not fully capture the diversity of patterns in localized prostate cancer and benign conditions, hindering their utility for slide-level Gleason assessments. The PANDA challenge \cite{bulten2022artificial} introduced a substantial dataset, albeit with biopsy-level Gleason score annotations, not aligning with the research's goals. Additionally, two datasets, the Gleason19 challenge and SICAPv2, are thoroughly discussed as follows:\\
The SICAPv2 database, with 155 biopsies from 95 consenting participants, addresses concerns of inter-observer variability through collaborative Gleason score assignments by Valencia hospital pathologists. Additionally, the Gleason19 challenge dataset, featuring 331 annotated cores from 244 individuals, provides meticulously prepared and annotated Prostate cancer TMA images, contributing valuable resources for prostate. 

\section{Methods}
\subsection{Data Preparation}
\paragraph{Gleason2019} The Gleason 2019 dataset comprises TMA images with a resolution of 5120 x 5120 pixels, which are annotated by several experts. After calculating the final annotation map for each image by pixel-level majority voting, we opt for smaller patches inside original images to train our model. Specifically, we extract 512 x 512 patches with a stride of 256 in each image, and assign a single label for them based on pixel annotations at their central areas. If all of the pixels residing in this 250 x 250 area do not belong to the background and share an identical label, the patch label is set accordingly, otherwise the patch is discarded.\\
After generating the patches, we use a patient-based cross-validation approach. This involves reserving 20\% of the data for final testing, while 10\% of the remaining training set is randomly chosen for validation during the training process.
\paragraph{SICAPv2} The SICAPv2 dataset offers whole slide histology images with both global Gleason scores and patch-level grades. Like the Gleason2019 dataset, the patches are 512 x 512 pixels with a 50\% overlap. We implemented a cross-validation strategy in which each patient and their associated patches are assigned to a single fold, resulting in four folds and a separate test subset. Notably, they meticulously selected these patients to guarantee the class balance between cross-validation subsets. In our experiments we have followed their method for preparing our training data.

\begin{figure}[h]
  \centering
    \includegraphics[width=0.8\columnwidth]{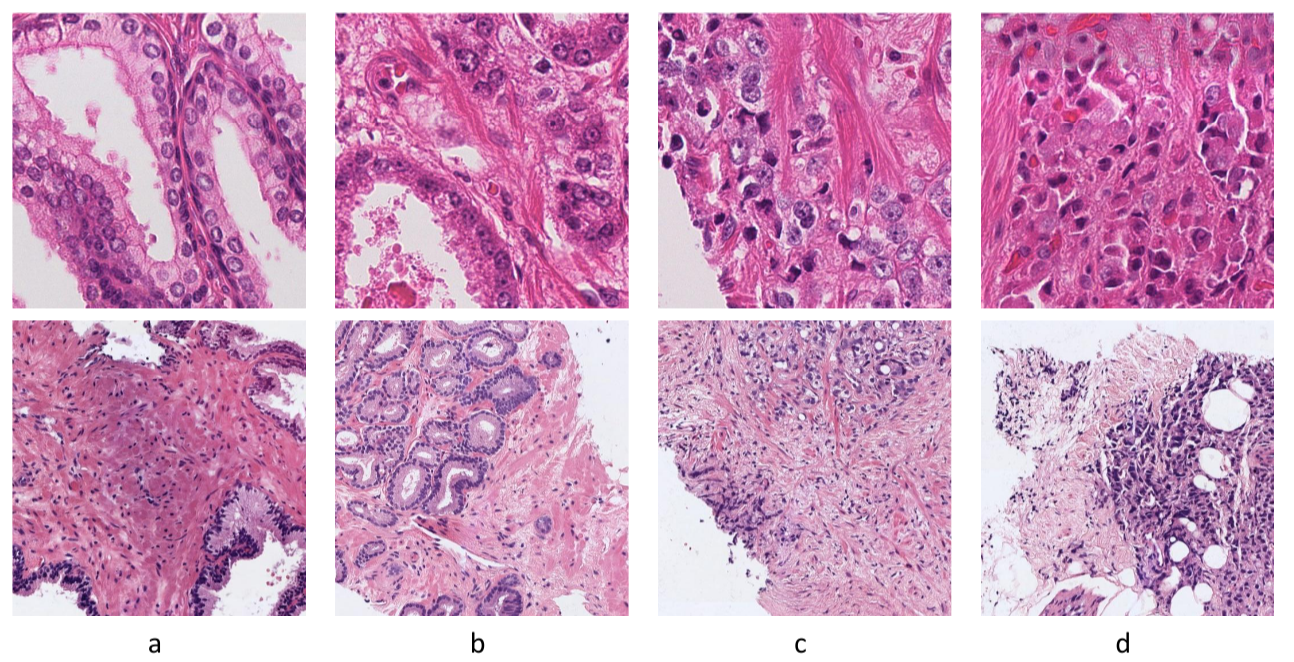}
    \caption{Examples of TMA images showcasing Gleason grades from the Gleason2019 dataset (top row) and the SICAPv2 dataset (bottom row): a) benign tissue, b) Gleason grade 3, c) Gleason grade 4, and d) Gleason grade 5.}
  \label{fig:fig1}
\end{figure}

Figure \ref{fig:fig1} illustrates some instances of TMA images in both datasets on 4 classes (i.e. benign, G3, G4, G5). Differentiating between these grades presents a significant challenge, even for seasoned pathologists, due to the subtle variations in cellular morphology and tissue architecture characteristic of each grade.

Table.\ref{table:num} presents the count of patches with regard to their Gleason grading categories for Gleason2019 and SICAPv2 datasets.

\begin{table}[h]
\centering
\caption{Number of patches corresponding to Gleason categories in Gleason2019 and SICAPv2 datasets.}
\begin{tabular}{lccccc}
\toprule
 & Benign & Grade 3 & Grade 4 & Grade 5 & Total \\
\midrule
Gleason2019 & 2767 & 10073 & 15502 & 312 & 28654 \\
SICAPv2 & 4417 & 1636 & 3622 & 665 & 10340 \\
\bottomrule
\end{tabular}
\label{table:num}
\end{table}

\subsection{YOLO and Its Application in Histopathology Images}
The YOLO (You Only Look Once) algorithm \cite{redmon2016you}, renowned for its efficiency and accuracy in real-time object detection, operates by dividing an input image into a grid. Each grid cell predicts a certain number of bounding boxes along with the confidence scores for those boxes, all within a single forward pass through the network. This approach significantly reduces the computational cost compared to traditional methods that involve multiple stages of detection and classification. YOLO's grid-based approach enables it to detect small objects effectively, and its use of anchor boxes allows the model to handle objects of varying scales within the same image. Moreover, YOLO is designed to efficiently utilize GPU resources, facilitating faster training and inference times.
\newline Despite its primary design for object detection and potential inaccuracies compared to other methods for specific tasks, YOLO's advantages in real-time detection and overall efficiency make it a compelling choice for applications in medical imaging, where rapid and accurate analysis of complex visual data is paramount. Given these strengths, including its speed, accuracy, and ability to handle multiple scales, YOLO was chosen as one of the three deep learning methods employed in this study for the classification of histopathology images.
\newline Salman's groundbreaking article represents a pivotal moment in integrating artificial intelligence (AI), specifically the YOLO algorithm, into the domain of digital pathology \cite{salman2022automated}. This trailblazing research aims to automate the detection and classification of prostate cancer in histopathology images, capitalizing on AI's advanced capabilities to boost diagnostic precision and efficiency. The study uniquely explores YOLO's applicability in analyzing Gleason histopathology images, marking a first-of-its-kind endeavor in academic literature. The research methodology is meticulously crafted around the assembly of a custom dataset, meticulously selected to balance quantity and quality. This dataset includes annotated prostate tissue biopsy images, meticulously annotated by pathologists to uphold the utmost accuracy. Through strategic data augmentation techniques, the dataset is enlarged, introducing a diversity of examples to mimic real-world conditions and bolster the model's learning capability. As the investigation unfolds, the YOLO model undergoes fine-tuning to discern the unique patterns linked to various Gleason grades, allowing it to distinguish between benign and malignant tissues with exceptional precision. This flexible strategy caters to the subtleties of histopathological image analysis, demonstrating YOLO's flexibility and versatility in managing intricate visual data. In summary, the adoption of YOLO for histopathology image classification in Salman's study leverages the algorithm's real-time detection capabilities, high accuracy, and scalability to tackle the complex task of prostate cancer diagnosis. Despite its limitations, particularly in detecting small objects and variations in lighting conditions, YOLO's strengths in speed and efficiency position it as a valuable tool in the field of digital pathology, contributing to the advancement of AI-driven diagnostics.
\newline In this study, we have employed YOLOv8x for the classification of Gleason grades in prostate cancer images. YOLOv8x, the most advanced variant in the YOLOv8 family developed by Ultralytics, offers state-of-the-art performance for image classification tasks. This model variant leverages sophisticated neural network architectures and training techniques to achieve the highest accuracy among YOLOv8 models. While computationally demanding, YOLOv8x's superior capabilities make it particularly well-suited for complex medical image analysis tasks such as Gleason grade classification, where precision is paramount. 

\subsection{Transformers in Histopathology Image Analysis}
Transformers have emerged as a pivotal technology in the realm of artificial intelligence, particularly in the analysis of medical images and their classification \cite{vafaeezadeh2023carpnet, vafaeezadeh2024ultrasound}. Initially designed for natural language processing (NLP) tasks, transformers have been ingeniously adapted for image classification, marking a significant advancement in the field of computer vision.
At the heart of transformers is the concept of transforming or changing an input series into an output series by acquiring contextual understanding and tracking relationships between series elements. This capability is particularly beneficial in medical imaging, where understanding the context and relationships within images is crucial for precise diagnosis and therapeutic planning.
Vision transformers (ViT) \cite{dosovitskiy2020image} represent a groundbreaking adaptation of the transformer paradigm for image classification tasks. By treating an image as a sequence of fixed-size patches rather than a grid of pixels, ViTs enable the model to detect associations between any pair of segments, irrespective of their location. This approach mimics the way words are treated in a sentence, enabling the model to understand the context and relationships within the image data. The addition of positional embeddings ensures that spatial information is maintained, further enhancing the model's capacity for accurate image classification.\\
The application of transformers in medical imaging offers numerous benefits. Their ability to process all elements of the input sequence in parallel, combined with their capacity to capture long-range dependencies, makes them exceptionally suited for analyzing complex medical images. This efficiency holds particular significance in healthcare settings, where rapid analysis of medical images can significantly impact patient outcomes.\\
Moreover, transformers' adaptability and scalability unveiled fresh pathways for research and innovation in medical imaging. Their success in tasks such as object detection, image classification, and even segmentation suggests a promising future for their integration into clinical workflows, potentially revolutionizing diagnostics and treatment strategies.
Vision transformers have found applications in the examination of histopathology images, particularly for the classification of Gleason cancer. A notable example is the development of a weakly-supervised machine learning model for prostate cancer assessment and Gleason grading of histopathology images \cite{behzadi2024weakly}. This model employs a novel approach that integrates transformers with other techniques to address the challenges of Gleason grading. Specifically, it uses transformers to extract discriminative areas in histopathology images through a Multiple Instance Learning (MIL) algorithm. Following this, a graph is constructed using these discriminative patches, and a Graph Convolutional Neural Network (GCN) is developed based on the gated attention mechanism to classify the image into its Gleason grades.\\
Furthermore, the application of transformers extends beyond Gleason grading to encompass the broader field of prostate cancer detection in ultrasound images \cite{harmanani2024benchmarking}. Researchers have explored the use of several transformer architectures, including Vision Transformers (ViT), Compact Convolutional Transformers (CCT), and Pyramid Vision Transformers (PvT), for detecting cancer in small regions of interest (ROI) within ultrasound images. These models are pretrained using VICReg and then fine-tuned with an additional MLP classifier to detect cancer in individual ROIs. The aggregated predictions from each patch within a core are averaged to produce the final output for the core. This approach highlights the versatility of transformers in processing medical image data, demonstrating their potential to improve the accuracy of prostate cancer detection from ultrasound images.\\
In this paper we applied the ViT\_base\_patch16\_224\_in21k as a pre-trained Vision Transformer model trained on the ImageNet21k dataset, featuring a 16x16 patch size and a 224x224 input size. ImageNet21k, unlike the standard ImageNet-1K, offers a broader array of images and classes, making it more comprehensive and varied. However, its complexity, limited accessibility, and underestimated benefits lead to its infrequent use for pretraining deep learning models for computer vision tasks.\\
The Vision Transformer's image classification process comprises dividing the image into fixed-size patches, linearly embedding each patch, adding position embeddings, encoding the sequence with a Transformer, and utilizing the encoded output for classification. Position embeddings, representing the spatial location of each patch, are incorporated into the patch embeddings, typically learned alongside other model parameters during training.\\
Linearly embedding a patch in Vision Transformer involves transforming the pixel values of the patch into a high-dimensional vector space using a fully connected layer with learnable weights and biases, resulting in the patch embedding.\\
The Vision Transformer's primary advantage over CNN models lies in its ability to train end-to-end on large datasets without the need for manually engineered features or data augmentation, thanks to its foundation in the Transformer, adept at processing sequential data. Moreover, it has achieved state-of-the-art performance on several benchmark image classification datasets, including ImageNet.

\subsection{Vision Mamba for State-Space Modeling in Histopathology Images}
\begin{figure}[h]
  \centering
    \includegraphics[width=0.7\columnwidth]{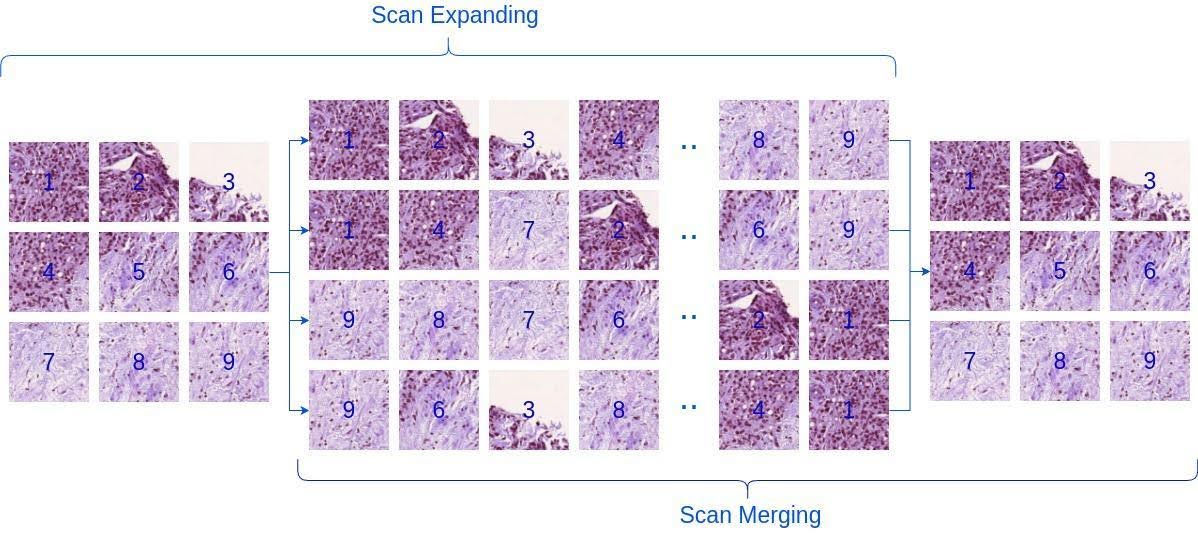}
    \caption{Main operations of 2D-selective-scan.}
  \label{fig:fig2}
\end{figure}

Vision Mamba is a novel approach for image classification, particularly tailored for medical images \cite{yue2024medmamba}. This method leverages the principles of Mamba, originally designed for large language models, to tackle the challenges of computer vision tasks, including image classification.\\
State Space Sequence Models (SSMs) are systems designed to map 1-dimensional sequences using a linear Ordinary Differential Equation (ODE). They offer benefits from previous architectures, such as parallel processing of data like transformers and linear complexity during inference similar to recurrent layers. However, their adoption has been limited due to issues like vanishing/exploding gradients and higher memory use compared to CNNs.\\
Structured State Space Sequence Models (S4) enhance basic SSMs by structuring the state matrix. This matrix is initialized with a High-Order Polynomial Projection Operator, enabling the creation of deep models with effective long-range reasoning. Mamba builds on SSMs for discrete data modeling (e.g., text and genome) by introducing two key innovations. First, it uses an input-dependent selection mechanism, unlike traditional time- and input-invariant SSMs, which improves information filtering by parameterizing SSM parameters based on input data. Second, Mamba incorporates a hardware-aware algorithm that scales linearly with sequence length, making it faster on modern hardware. The Mamba architecture combines SSM blocks with linear layers for a simpler design and has achieved state-of-the-art performance in long-sequence tasks, such as language and genomics, due to its efficiency in both training and inference.\\
The SS2D comprises three components: a scan expanding operation, an S6 block, and a scan merging operation. As depicted in Figure \ref{fig:fig2}, the scan expanding operation unfolds the input image into sequences along four directions (top-left to bottom-right, bottom-right to top-left, top-right to bottom-left, and bottom-left to top-right). These sequences undergo feature extraction in the S6 block, ensuring comprehensive scanning and capturing diverse features from various directions. Afterwards, the scan merging operation sums and merges the sequences from all four directions, restoring the output image to its original size.

The S6 block, inspired by Mamba \cite{gu2023mamba}, incorporates a selective mechanism atop S4, adapting the SSM's parameters based on the input. This allows the model to differentiate and retain relevant information while filtering out the irrelevant. The pseudo-code for the S6 block, given an input x with batch size b, token length l, and dimension d, can be outlined as follows:

\begin{equation}
\begin{aligned}
\Delta, B, C &= \text{Linear}(x), \text{Linear}(x), \text{Linear}(x) \\
A &= \exp(\Delta A) \\
B &= (\Delta A)^{-1} (\exp(\Delta A) - I) \cdot \Delta B \\
\hat{h}_t &= A h_{t-1} + B x_t \\
y_t &= C h_t + D x_t \\
y &= [y_1', y_2', \ldots, y_t']
\end{aligned}
\end{equation}

Where A, B, C, and D are learnable parameters, Linear(.) denotes the linear projection layer,  and y is the output feature map with the same shape as input.
The MedMamba model is a deep learning architecture designed for medical image classification tasks. It is inspired by the VMamba architecture \cite{Liu2024VMambaVS} and incorporates a novel hybrid basic block called SS-Conv-SSM, which stands for State Space Convolution - State Space Model. This block is the core element of the MedMamba model and is designed to efficiently handle both local feature extraction and the capture of long-range dependencies within medical images.

\begin{figure}[t]
  \centering
    \includegraphics[width=0.88\columnwidth]{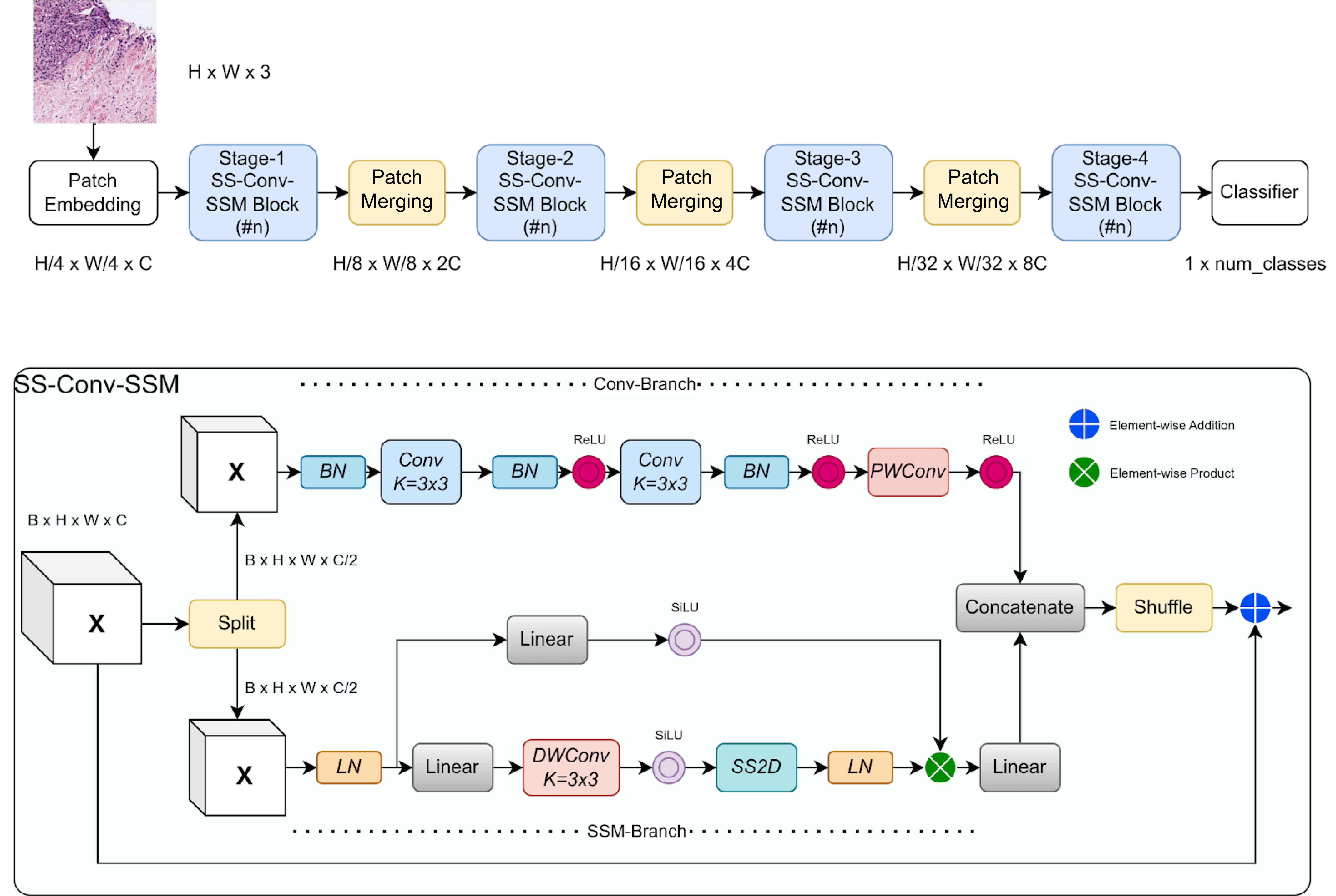}
    \caption{The architecture of MedMamba, cComprising Batch Normalization (BN), Layer Normalization (LN), Sigmoid Linear Unit (SiLU), Rectified Linear Unit (ReLU), linear layers, Point-wise Convolutions (PWConv), and Depth-wise Convolutions (DWConv). SS2D: Selective State Space Model 2D, B: Batch size, H: Height dimension, W: Width dimension, C: Channel dimension.}
  \label{fig:fig3}
\end{figure}

As depicted in Figure \ref{fig:fig3}, Overall architecture of MedMamba includes several key components:
Patch Embedding Layer: This layer is responsible for transforming the input medical images into a format suitable for processing by the subsequent layers of the model. It typically involves dividing the image into patches and embedding them into a lower-dimensional space.
Stacked SS-Conv-SSM Blocks: These are the building blocks of the MedMamba model. Each SS-Conv-SSM block consists of the following:\\\
\textbf{Channel-Split}: Divides the input channels into groups.\\
\textbf{Convolutional Layers:} Extract local features from the input.\\
\textbf{SSM Layers:} Capture long-range dependencies using state space models.\\
\textbf{Channel-Shuffle:} Re-arranges the channels to mix the features extracted by different groups.\\
\textbf{Patch Merging Layers: }These layers are used for down-sampling the feature maps, which helps in reducing the computational complexity and capturing hierarchical features at different scales.\\
\textbf{Feature Classifier: }The final component of the MedMamba model is a classifier that takes the high-level features extracted by the previous layers and uses them to classify the input image into one of the predefined categories.\\
The MedMamba model also employs a grouped convolution strategy and channel-shuffle operation to reduce the number of parameters and lower the computational burden, making it efficient for medical artificial intelligence applications. This design allows MedMamba to maintain excellent performance while being computationally efficient, addressing the limitations of both CNNs and ViTs in the context of medical image classification.\\
Vision Mamba establishes a new baseline for medical image classification, providing valuable insights for developing more powerful SSM-based artificial intelligence algorithms and application systems in the medical field. To conduct a fair comparison, we have employed Mamba and other methods based on their reported configurations and trained them on our datasets.\\

\section{Results}
We employ several classification metrics for quantitative evaluation of competing methods. These metrics are defined as follows:

\begin{equation}
\text{Precision} = \frac{\text{TP}}{\text{TP} + \text{FP}}
\end{equation}

\begin{equation}
\text{Recall} = \frac{\text{TP}}{\text{TP} + \text{FN}}
\end{equation}

\begin{equation}
\text{F-Score} = \frac{2 \times (\text{Precision} \times \text{Recall})}{\text{Precision} + \text{Recall}}
\end{equation}

\begin{equation}
\text{Accuracy} = \frac{\text{TP} + \text{TN}}{\text{TP} + \text{TN} + \text{FP} + \text{FN}}
\end{equation}

With TP, FP, TN, and FN indicating true positives, false positives, true negatives, and false negatives, respectively.\\
These metrics highlight the challenge in differentiating between four classes in Gleason2019 and SICAPv2 dataset. It is worth mentioning that due to severe class imbalance in these datasets, we opt for a weighted average over class metrics to provide a more comprehensive measurement of overall performance. This approach ensures that the performance metrics reflect the proportion of each class in the dataset, rather than being skewed by the performance on less frequent classes.\\
To compare the performance of aforementioned models, we have employed two public datasets and reported the results in Table.\ref{tabel:gl} and Table.\ref{tabel:si}. It can be seen that MedMamba outperforms other competing methods with regards to all quantitative metrics. Specifically, by achieving a high precision and recall, Medmamba showcases its ability in identifying True Positives while maintaining a low False Positive and False Negative rate. Compared to that, ViT presents inferior metrics with much higher computational complexity. Additionally, due to high resolution of histopathology images, long-range feature extraction plays a crucial role in overall model performance. Compared to CNNs, Mamba and ViT architectures excel in this regard and deliver better results.

\begin{table}[h!]
\centering
\caption{Results of competing methods on Gleason2019 dataset based on weighted metrics.}
\begin{tabular}{lcccc}
\toprule
 & Precision & Recall & F1-score & Accuracy \\
\midrule
Mamba              & 85.82  & 84.87  & 85.34  & 85.13  \\
YOLO               & 83.56  & 82.24  & 82.89  & 83.69  \\
Vision Transformer & 83.84  & 83.16  & 83.49  & 84.07  \\
\bottomrule
\end{tabular}
\caption{ Results of competing methods on Gleason2019 dataset based on weighted metrics.}
\label{tabel:gl}
\end{table}

\begin{table}[h!]
\centering
\caption{Results of competing methods on SICAPv2 dataset based on weighted metrics.}
\begin{tabular}{lcccc}
\toprule
 & Precision & Recall & F1-score & Accuracy \\
\midrule
Mamba              & 64.61  & 58.23  & 61.25  & 69.98  \\
YOLO               & 56.96  & 53.58  & 55.21  & 64.7  \\
Vision Transformer & 62.26  & 58.86  & 60.51  & 66.02  \\
\bottomrule
\end{tabular}
\label{tabel:si}
\end{table}

\begin{figure}[h]
  \centering
    \includegraphics[width=0.7\columnwidth]{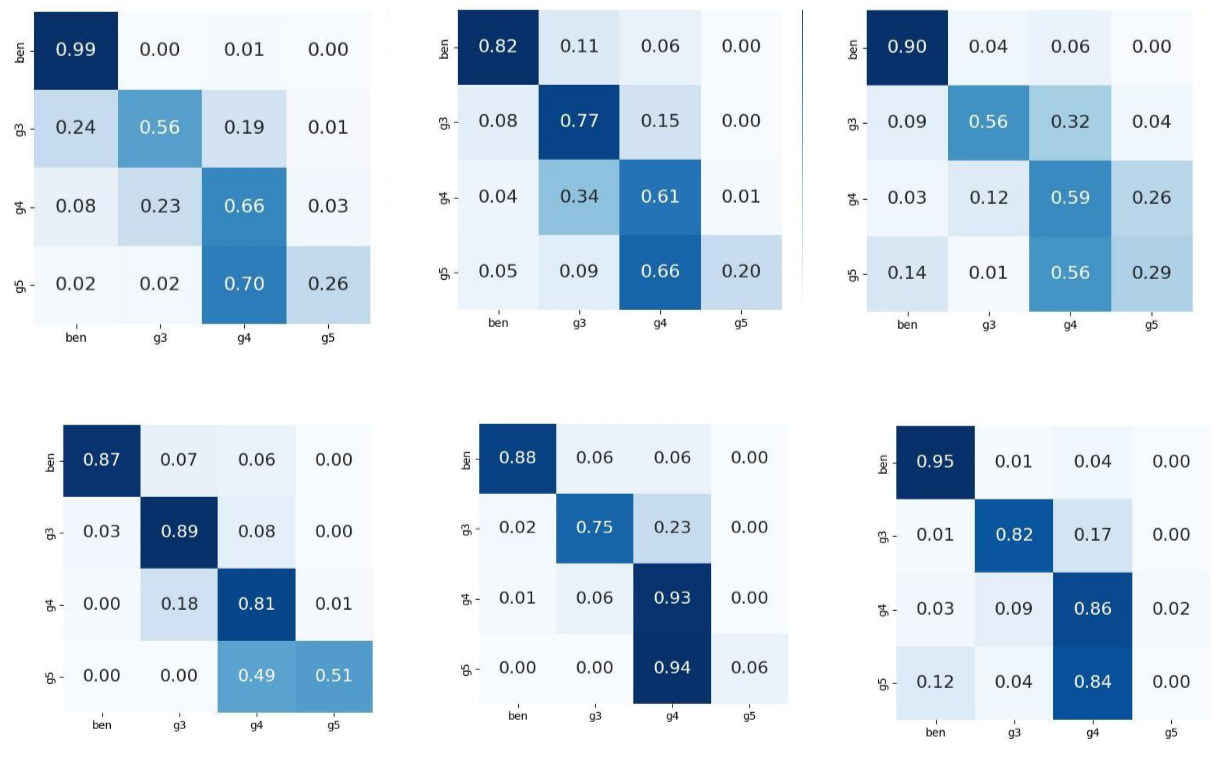}
    \caption{ Normalized confusion matrices on SICAPv2 (top) and Gleason2019 (bottom) datasets. Matrices from left to right correspond to Mamba, ViT, and YOLO models. Ben: Benign, g3: Grade 3, g4: Grade 4, g5: Grade 5.}
  \label{fig:fig4}
\end{figure}

Figure \ref{fig:fig4} represents the normalized confusion matrices for competing methods on two datasets, with the first and second row corresponding to Gleason2019 and SICAPv2 datasets, respectively.  These tables provide a comprehensive visual representation of models’ performance. It is evident that the diagonal of the tables, which represent the True positive predictions are lower for SICAPv2 dataset compared to Gleason, which can be attributed to its challenging samples. Overall, Mamba performs strongly on benign cases and well on the others, though it consistently struggles with higher Gleason grades, particularly g5. On the Gleason dataset, ViT encounters more difficulty, especially with g4 and g5, showing greater confusion between these classes than Mamba. YOLO surpasses ViT in handling benign cases, but still encounters challenges with g4 and g5. Additionally, it excels at distinguishing higher grades, particularly g5, in the SICAPv2 dataset.

\section{Discussion}
In recent years, the field of prostate cancer diagnosis and prognosis has witnessed significant advancements through the integration of artificial intelligence (AI) technologies. Researchers have developed novel AI-based methods that not only enhance the accuracy of diagnosis but also improve the prediction of patient outcomes.\\
One of the key breakthroughs in AI-based methods for prostate cancer diagnosis is the achievement of expert-level performance in lesion detection and Gleason grading. By leveraging deep learning algorithms, these methods have demonstrated capabilities comparable to human pathologists in accurately identifying lesions and classifying tissue samples based on Gleason patterns.\\
In the context of histopathology image analysis, YOLO's speed and efficiency are particularly advantageous. The rapid turnaround times enabled by YOLO can facilitate immediate feedback during surgical procedures or expedite the triage of biopsy samples, potentially leading to quicker diagnosis and treatment initiation. Moreover, YOLO's architecture allows for the simultaneous detection of multiple objects within an image, a feature that is highly beneficial in histopathology where multiple types of cells and tissues may be present in a single slide.\\
However, while YOLO's speed and flexibility offer significant advantages, the model's performance in histopathology image classification is contingent upon careful tuning and validation. The nuances of histopathological images, characterized by subtle variations in cellular morphology and tissue composition, pose unique challenges that necessitate the optimization of YOLO's parameters and the selection of appropriate training datasets. Furthermore, the balance between speed and accuracy remains a critical consideration, as overly aggressive optimizations aimed at increasing speed may compromise the model's ability to accurately distinguish between different Gleason grades or other pathological features.\\
Despite these challenges, the integration of YOLO into the pipeline for histopathology image analysis represents a promising direction for advancing diagnostic workflows. By leveraging YOLO's strengths in real-time object detection, researchers and clinicians can explore new paradigms for integrating AI into clinical practice, potentially revolutionizing the way prostate cancer is diagnosed and managed.\\
CNNs and ViTs both have limitations when it comes to medical image classification tasks. CNNs are adept at extracting local features due to their convolutional layers, but they struggle with capturing global context and long-range dependencies because of their inherently limited local receptive fields. This limitation can lead to insufficient feature extraction and suboptimal classification results.\\
On the other hand, ViTs, which were originally designed for natural language processing, represent an input image as a sequence of image patches and use self-attention mechanisms to capture long-range dependencies effectively. However, ViTs lack the ability to handle spatial image hierarchies natively, and their self-attention mechanism has a high quadratic complexity, which can degrade local feature details. This complexity results in a significant computational burden, especially for high-resolution medical images, making it challenging to deploy ViT models in clinical settings with limited computational resources.\\
MedMamba integrates the advantages of CNNs and SSMs for efficient medical image classification by introducing a novel hybrid basic block called SS-Conv-SSM. This block combines the capabilities of convolutional layers to extract local features with the strengths of SSMs in capturing long-range dependencies. The SS-Conv-SSM block is designed to efficiently process medical images by using a combination of channel-split, convolutional layers, SSM layers, and channel-shuffle operations.\\
The convolutional layers in the SS-Conv-SSM block are responsible for capturing local spatial features from the input images, which is a task that CNNs are well-suited for due to their ability to learn hierarchical representations of visual data. On the other hand, the SSM layers within the block are adept at modeling long-range dependencies, which is crucial for understanding the global context of medical images. By stacking multiple SS-Conv-SSM blocks, MedMamba can build a deep architecture that effectively captures both local and global features from medical images of different modalities.\\
Furthermore, MedMamba employs a grouped convolution strategy and channel-shuffle operation to reduce the number of parameters and lower the computational burden, making it efficient for medical artificial intelligence applications. This design allows MedMamba to maintain excellent performance while being computationally efficient, addressing the limitations of both CNNs and ViTs in the context of medical image classification.\\

\section{Conclusion}
This study has comprehensively evaluated and compared the performance of three deep learning methodologies(i.e.,YOLO, Vision Transformers, and Vision Mamba) in the classification of Gleason grades from histopathology images. The goal was to enhance diagnostic precision and efficiency in prostate cancer management, a critical issue given the prevalence of prostate cancer and the reliance on Gleason grading for diagnosis and prognosis.\\
The results demonstrate that Vision Mamba emerged as the most effective model for Gleason grade classification, offering a balance between accuracy and computational efficiency. It achieved high precision and recall rates while minimizing false positives and negatives, showcasing its superiority across all evaluated metrics. YOLO, despite its speed and efficiency advantages, particularly in real-time analysis, was found to be less accurate in this specific task. Vision Transformers, while capable of capturing long-range dependencies within images, exhibited higher computational complexity compared to the other models.\\
These findings suggest that Vision Mamba could significantly enhance the precision of prostate cancer diagnosis and treatment planning by integrating into diagnostic workflows. Its ability to accurately classify Gleason grades from histopathology images could streamline the diagnostic process, potentially leading to earlier and more accurate diagnoses, improved patient outcomes, and reduced healthcare costs.\\
Further research is recommended to optimize model parameters and explore the applicability of these deep learning techniques in broader clinical contexts. Such efforts could pave the way for the widespread adoption of AI in pathology, revolutionizing the field and improving patient care. The availability of the code used in this study on GitHub (https://github.com/Swiman/mamba-medical-classification) encourages further exploration and experimentation by the scientific community, fostering innovation in the application of AI to medical imaging.\\
In conclusion, the integration of deep learning techniques, particularly Vision Mamba, into the diagnostic workflow for prostate cancer holds great promise. It not only enhances the accuracy and efficiency of Gleason grading but also underscores the potential of AI to transform healthcare delivery. As the field continues to evolve, it will be essential to continue investigating and refining these methodologies to ensure they remain at the forefront of diagnostic innovation.


\bibliographystyle{unsrt}  
\bibliography{references}

\end{document}